\documentclass[11pt]{article}

\usepackage{geometry}
\usepackage[utf8]{inputenc}
\usepackage{textcomp}
\usepackage[backend=biber, 
style=phys, 
pageranges=false, 
language=english, 
biblabel=brackets
]{biblatex}
\usepackage{graphicx}
\usepackage{booktabs}
\usepackage{authblk}
\usepackage{siunitx}
\usepackage{upgreek}
\usepackage{amsmath}
\let\Lambda\varLambda

\usepackage{titlesec}
\titleformat*{\section}{\large\bfseries}
\titleformat*{\subsection}{\normalsize\bfseries}
\titleformat*{\subsubsection}{\large\bfseries}
\titleformat*{\paragraph}{\large\bfseries}
\titleformat*{\subparagraph}{\large\bfseries}

\title{Parameter Optimization of Light Outcoupling Structures for High-Efficiency Organic Light-Emitting Diodes}

\author[1]{Dinara Samigullina}
\author[1]{Paul-Anton Will}
\author[2]{Lydia Galle}
\author[1]{Simone Lenk}
\author[2]{Julia Grothe}
\author[2]{Stefan Kaskel}
\author[1,*]{Sebastian Reineke}

\affil[1]{Dresden Integrated Center for Applied Physics and Photonic Materials (IAPP) and Institute for Applied Physics, Technische Universität Dresden, Nöthnitzer Str. 61, 01187 Dresden, Germany}
\affil[2]{Inorganic Chemistry I, Technische Universität Dresden, Bergstrasse 66, 01069 Dresden, Germany}
\affil[*]{sebastian.reineke@tu-dresden.de}

\date{\today}

\begin{document}

\maketitle

\textit{[The following article has been submitted to Journal of Applied Physics.]}\\

Organic light-emitting diodes (OLEDs) have successfully entered the display market and continue to be attractive for many other applications. As state-of-the-art OLEDs can reach an internal quantum efficiency (IQE) of almost \SI{100}{\%}, light outcoupling remains one of the major screws left to be turned. The fact that no superior outcoupling structure has been found underlines that further investigations are needed to understand their prospect. In this paper, we use two-dimensional titanium dioxide (2D TiO\textsubscript{2}) block arrays as a model of an internal light outcoupling structure and investigate the influence of its geometrical parameters on achieving the highest external quantum efficiency (EQE) for OLEDs. The multivariable problem is evaluated with the visual assistance of scatter plots, which enables us to propose an optimal period range and block width-to-distance ratio. The highest EQE achieved is \SI{45.2}{\%} with internal and external structures. This work contributes to the highly desired prediction of ideal light outcoupling structures in the future.

\section{Introduction}

Organic light-emitting diodes (OLEDs) are an innovative technology for backlight-free displays and can nowadays be found in diverse consumer electronic devices such as smartphones, TVs, and smartwatches. 
Besides their attractiveness for automotive and lighting applications, OLEDs have a high potential to be used as biological, health-monitoring sensors and wearable electronics \cite{heoRecentProgressTextileBased2018, songOrganicLightEmittingDiodes}. However, to enable a successful transition to these new markets, further improvements of OLED stability and efficiency need to be achieved.

The efficiency of OLEDs is expressed with the external quantum efficiency (EQE), which is defined by the product of the internal quantum efficiency (IQE) times the light outcoupling efficiency. The first factor can achieve \SI{100}{\%} by using highly efficient emitters, such as phosphorescent \cite{100phosphorescent, baldoHighlyEfficientPhosphorescent1998, adachiNearly100Internal2001} and thermally activated delayed fluorescent emitters \cite{zhangEfficientBlueOrganic2014, goushiOrganicLightemittingDiodes2012, uoyamaHighlyEfficientOrganic2012}, and by engineering an electrically efficient stack of organic layers with transport and blocking layers \cite{walzerHighlyEfficientOrganic2007a, ikaiHighlyEfficientPhosphorescence2001}. The EQE is therefore mostly limited by the extraction of the generated photons, which are confined due to the high refractive indices of the organic layers and substrate. This planar geometry leads to formation of various lossy modes, i.e. waveguided modes in organic layers and the substrate \cite{bruttingDeviceEfficiencyOrganic2013}. Additionally, a fraction of light is absorbed by the OLED materials and coupled to surface plasmon polaritons at metal electrodes \cite{fuchsEnhancedLightEmission2015}.  

Theoretically, a planar OLED can reach EQEs up to \SI{46}{\%} with a perfect IQE and horizontally oriented  emitters \cite{KimEQE46}. Practically, neither the emitters nor the IQE show ideal properties, which is why the EQE is typically limited to 20--30\,$\%$. To exceed this natural EQE limit, the waveguided and substrate modes need to be redirected to the air with light outcoupling structures, while additionally, surface plasmon polariton modes at metal interfaces must be suppressed with carefully selected refractive indices~\cite{songLensfreeOLEDs502018, fuchsEnhancedLightEmission2015}. For the former, so-called internal and external light outcoupling structures are used for OLEDs. The internal structures are implemented into the active layers of the OLED stack and target waveguided and surface plasmon polariton modes. The external structures aim for substrate modes and are applied to the outside of the OLED. There is a large variety of internal structure examples in literature. Most of them are based on  scattering the light, e.g nanohole arrays \cite{JeonNanoholeArray}, periodically corrugated substrates \cite{MattersonCorrugatedSub, SchwabCorrugated, younCorrugatedSapphireSubstrates2015, CHO2017139}, 2D photonic crystals \cite{Do2Dcrystal, Lee2Dcrystal, ishiharaOrganicLightemittingDiodes2007}, nanoparticle scattering layers \cite{ChangNanoparticle, CHANG20121073, ChangNanoparticle2, preinfalkLargeAreaScreenPrintedInternal2017} and low-index grids \cite{sunEnhancedLightOutcoupling2008}. While there is no superior internal structure, the most efficient external light outcoupling structure is a glass half-sphere. It can simply be attached to the substrate and can increase the EQE by more than a factor of two \cite{leeSynergeticElectrodeArchitecture2016, reinekeWhiteOrganicLightemitting2013, leeHighEfficiencyOrangeTandem2014}. However, the bulky half-sphere is not suitable for large-area applications, which is why microlens arrays were introduced \cite{mollerImprovedLightOutcoupling2002, stellmacherFastCostEffective2019, madiganImprovementOutputCoupling2000}. 

The light outcoupling structures listed above possess various parameters such as arrangement, period, thickness, chemical composition (e.g. concentration of nanoparticles), and refractive index. All these parameters need to be controlled in order to achieve the highest light output. However, it becomes difficult to find a general recipe for the highest EQE when combining the light outcoupling structure with the OLED. The reason is the influence of many different structure parameters on the OLED's functionality and its efficiency in various ways. For example, how the period or height influences light extraction and what the best combination for the highest EQE is. 

In this paper, we investigated periodic two-dimensional titanium dioxide (2D TiO\textsubscript{2}) block arrays that are used as internal light outcoupling structures for OLEDs. The period, height, and distance between the blocks were varied to study their influence on the EQE. In addition, a glass half-sphere is used as external light outcoupling structure in combination with the TiO\textsubscript{2} blocks to further increase the EQE. While previous works mostly focused on fabrication of light outcoupling structures and its implementation into the OLEDs, we aim to find the combination of structure parameters which leads to the highest EQE. For this purpose, we systematically evaluated the complex multi-dimensional combination of parameters with the assistance of scatter plots. With this, we are able to propose a range for the ideal period and an ideal block width-to-distance ratio. With the best combination of parameters, we can enhance the EQE of the reference OLED from \SI{20.3}{\%} to \SI{45.2}{\%} with implementing the internal structures and external together.

Firstly, we describe the fabrication and characterization of the 2D TiO\textsubscript{2}  blocks. Then we present the employed OLEDs with their fabrication, optoelectrical characteristics, EQEs, and angular resolved spectra. Afterwards, we analyse the respective EQE enhancement and correlate it to the block parameters.  

\section{Experimental methods}

\subsection{Fabrication and characterization of 2D TiO\textsubscript{2} blocks as internal light outcoupling structures}

A TiO\textsubscript{2} precursor solution is prepared by following the procedure of Richmond \textit{et al.} \cite{richmond2011pressureless}. 
A microcontact printing tool ($\upmu$ContactPrinter 3.0, GeSiM mbH) is used to fabricate 2D TiO\textsubscript{2} block arrays. A 6\,×\,6" glass plate (Eagle XG, Corning) is used as a substrate. It consists of 36 indium tin oxide (ITO) layouts (Thin Film Devices Inc.), which are pre-structured with four adjacent electrodes with a height of \SI{90}{\nm}. A drop of the precursor solution is deposited on top of two electrodes of the substrate. A polydimethylsiloxane (PDMS) stamp is pressed into the precursor drop with constant pressure while the substrate is heated to \SI{50}{\degreeCelsius}. The stamp has a negative profile of the desired structures. Afterwards, the printed structures are placed in an oven and heated to \SI{600}{\degreeCelsius} for two hours with a heating ramp of 1\,K per min. Then, they are treated in pure argon atmosphere at \SI{600}{\degreeCelsius} for five hours with a ramp of 5\,K per minute to regain the ITO conductivity. More detailed information on the printing technique can be found in Reference~\cite{wisser2015precursor}.

The produced block arrays are characterized by atomic force microscopy (AFM) with a Combiscope (AIST-NT) and scanning electron microscopy (SEM) (ZEISS DSM-982 Gemini). The transmission is measured by a UV-VIS-NIR Spectrophotometer (SolidSpec-3700, Shimadzu Corp.).

\subsection{Fabrication and characterization of OLEDs}

The substrate with printed TiO\textsubscript{2} block arrays is cleaned before the deposition of organic layers by the procedure that follows. The substrate is cleaned with acetone and ethanol using a cotton tissue, then it undergoes an ultrasonic treatment in n-methyl-2-pyrrolidone (NMP) for 15 minutes and is rinsed with distilled water. Following this, it is put again in an ultrasonic bath, firstly filled with distilled water for 10 minutes and then with ethanol for 15 minutes and dried under a nitrogen stream. Finally, it undergoes oxygen plasma treatment for 10 minutes. The OLEDs are fabricated in a single evaporation chamber (Kurt J. Lesker Co.) under an ultra-high vacuum ($10^{-8}$ - $10^{-7}$\,mbar). The layer thicknesses and evaporation rates are monitored by quartz crystal microbalances. All samples are produced simultaneously on the ITO layouts. Two ITO electrodes on each sample are structured with 2D TiO\textsubscript{2} block arrays and two remain planar for reference purpose. The OLED stack (Figure 1d) consists of \SI{56}{\nm} N,N'-((diphenyl-N,N'-bis)9,9,-dimethyl-fluoren-2-yl)-benzidine (BF-DPB) doped with \SI{4}{wt\%} 2,2'-(perfluoronaphthalene-2,6-diylidene)dimalono-nitrile (F\textsubscript{6}-TCNNQ), \SI{10}{\nm} N,N'-di(naphthalene-2-yl)-N,N'diphenyl-benzidine (NPB), 20~nm NPB doped with \SI{10}{wt\%} iridium(III)bis(2-methyldibenzo[f,h]quinoxaline)(acetylacetonate) (Ir(MDQ)\textsubscript{2}(acac)), \SI{10}{\nm} bis-(2-methyl-8-chinolinolato)-(4-phenyl-phenolato)-aluminium(III) (BAlq\textsubscript{2}), \SI{74}{\nm} of  4,7-diphenyl-1,10-phenanthroline (BPhen) doped 1:1 with cesium (Cs) and \SI{100}{\nm} silver (Ag). After fabrication, all OLEDs are immediately encapsulated under nitrogen atmosphere with a glass lid.

The electrical characteristics and efficiencies of all OLEDs are obtained in an integrating sphere (LMS-100, Labsphere Inc.). The emission spectra are measured at a current density of $j = 15$\,mA/cm$^2$ (Keithley SMU2400). A glass half-sphere with a diameter of \SI{10}{\milli\meter} was attached to the devices with an index-matching oil (Zeiss, Immersol 518F, $n = 1.52$) to extract substrate modes. The OLED pixel size is $\approx$\,2.5\,×\,2.5~mm. The angular resolved emission spectra are measured at $j = 15$\,mA/cm$^2$ using a spectrometer (USB-4000, Ocean Optics) with 1$^\circ$ steps.

\section{Results}

\subsection{Design of an OLED with internal light outcoupling structures}

\begin{figure}[h!]
	\centering
	\includegraphics[scale = 0.65]{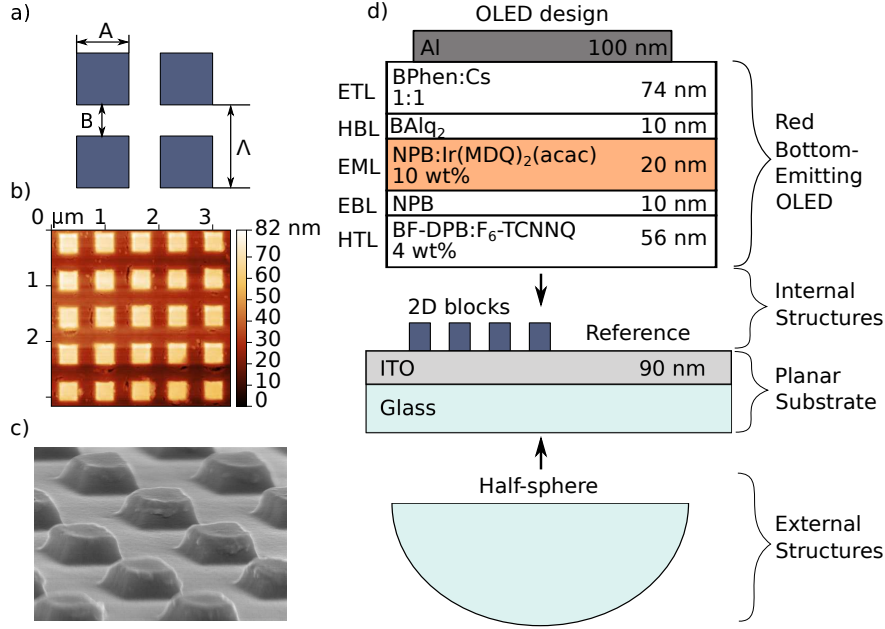}
	\caption{The investigated OLED with its light outcoupling structures. a) A sketch of the TiO\textsubscript{2} block array that is characterized by the period $\Lambda = A + B$, where $A$ is the block width and $B$ the distance between two blocks. b) Atomic force microscopy (AFM) image of a 2D TiO\textsubscript{2} block array with a period of 1~$\upmu$m. c)~Scanning electron microscope (SEM) image of a block array. d) Bottom-emitting OLED with structured and planar reference substrates. Additionally, a half-sphere is applied to extract substrate modes. ETL -- electron transport layer, HBL -- hole blocking layer, EML -- emission layer, EBL -- electron blocking layer, HTL -- hole transport layer.}
	\label{fig:blocks-oled}
\end{figure}

Figure 1 shows the investigated OLED architecture and how the light outcoupling structures were integrated. A periodic 2D block arrays serve as an internal light outcoupling structure to extract waveguided modes. Additionally, a glass half-sphere is used as an external light outcoupling structure that extracts substrate modes. A schematic sketch of crucial block array parameters is shown in Figure 1a. To study the impact on the OLED efficiency, the block array parameters are varied as described in the following.

We chose TiO\textsubscript{2} as a material for the blocks due to its high refractive index, close to $n=2.4$ \cite{richmond2011pressureless}, and straightforward fabrication \cite{wisser2015precursor}. The refractive index provides enough contrast to the organic layers ($n\approx1.7$~\cite{luOptimizationExternalCoupling2001}) to act as an optical grating. Furthermore, the nanoimprinting of TiO\textsubscript{2} is fast and easy and allows up-scaling to larger areas via roll-to-roll printing techniques \cite{kooyReviewRolltorollNanoimprint2014}. The stamp defines the period $\Lambda$ and the lateral dimensions $A$ and $B$. Figure 1b and Figure 1c show example AFM and SEM images from which the block parameters $\Lambda$, $A$, $B$, and height~$H$ were extracted. Note that the height $H$ of the blocks is difficult to control. First, the precursor, filling the stamp by capillary forces, does not fully penetrate into the wells. Second, the blocks shrink during the post-treatment process in the oven such that the final set of samples has some deviation in the parameters. This results in different heights for each period. All parameters of the investigated block arrays are listed in Table 1. A total of 8 samples (G1-G8) with different parameters were produced for further analysis. 

Figure 1d shows the state-of-the-art OLED layer sequence that is fabricated on top of the substrate with planar ITO as well as ITO with 2D block arrays. The OLED stack is optimized to achieve the highest possible EQE before applying outcoupling structures. For this purpose, we chose the red phosphorescent emitter Ir(MDQ)$_2$(acac), which has an anisotropic coefficient of $a = 0.25$ \cite{EmitterOrientation}. Therefore, it has a preferred horizontal orientation of emitting dipoles with respect to the OLED plane to ensure maximum light emission towards the substrate. Thicknesses of the transport layers are tuned to enhance the cavity effect following the work of P.-A. Will \textit{et al.}~\cite{will2019efficiency}. The reference and structured pixels are located next to each other on the same substrate to ensure an accurate comparison of OLED performance, which is discussed in the next section.

\begin{table}[h]
	\centering
	\caption{Parameters of the 2D TiO\textsubscript{2} block arrays and corresponding OLED EQEs. $\Lambda$ -- period, $A$ -- block width, $B$ -- distance between the blocks, $H$ -- height, $T$ -- transmission, EQE\textsubscript{bl} and EQE\textsubscript{hs} -- external quantum efficiencies (EQEs) of samples with block arrays and attached half-sphere at the current density $j$~=~15\,mA/cm$^2$.}
	\begin{tabular}{  c  c  c  c c c c c} 
		\toprule
		Name & 
		$\Lambda$, nm & A, nm & B, nm & H, nm & T, \% & EQE\textsubscript{bl}, \% & EQE\textsubscript{hs}, \%\\
		\midrule
		G1 & 
		600 & 400 & 200 & 117 & 88.32 & 20.6 & 40.1 \\
		
		G2 & 
		700 & 500 & 200 & 90 & 86.99 & 21.8 & 41.0 \\
		
		G3 & 
		800 & 500 & 300 & 21 & 89.55 & 20.8 & 42.0 \\
		
		G4 & 
		1000 & 500 & 500 & 105 & 87.86 & 21.4 & 42.9 \\
		
		G5 & 
		1500 & 750 & 750 & 90 & 87.0 & 20.7 & 43.0 \\
		
		G6 & 
		1200 & 1000 & 200 & 123 & 87.0 & 19.6 & 40.6 \\
		
		G7 & 
		1500 & 1000 & 500 & 42 & 87.76 & 21.1 & 41.7 \\
		
		G8 & 
		2000  & 1000 & 1000& 170 & 86.9 & 20.6 & 41.8 \\ 
		\midrule
		Planar & -- & -- & -- & -- & -- & 19.1 & 35.2  \\
		\bottomrule
	\end{tabular}
	\label{table:blocks}
\end{table}

\subsection{OLED performance}

\begin{figure}[h!]
	\centering
	\includegraphics[scale = 0.75]{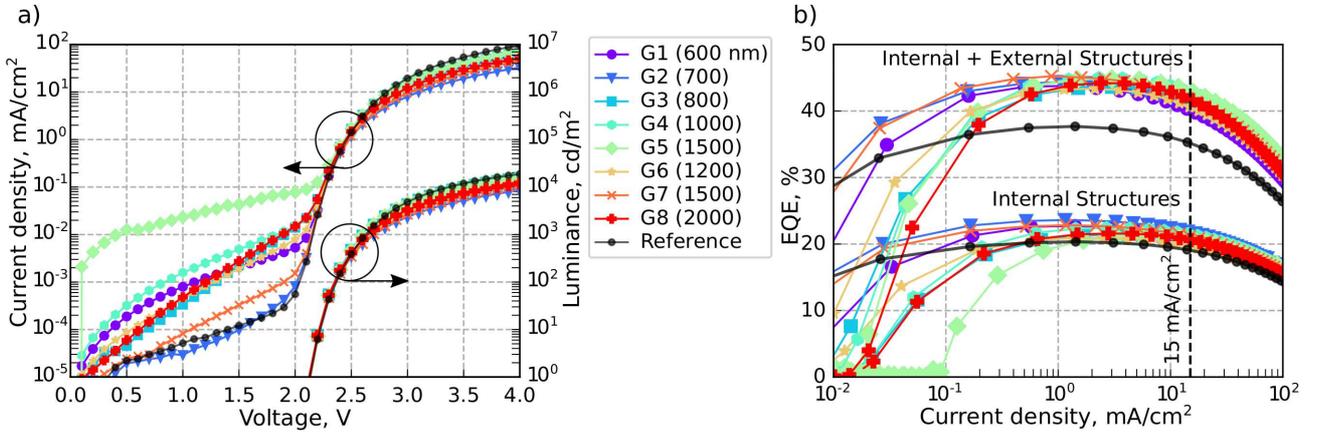}
	\caption{a) Current density-voltage-luminance (\textit{jVL}) characteristics of the structured OLEDs with different periods of block arrays and the planar reference. b) EQE as a function of current density of the structured OLEDs with different periods of block arrays.}
	\label{fig:IVL-EQE}
\end{figure}

Figure 2 presents the optoelectrical characteristics of the internally structured devices with varied periods of the block arrays and the planar reference. The current density plotted versus voltage reveals that the blocks alter the electrical performance of the OLEDs. At voltages $V < 2$\,V, the OLEDs with blocks show higher leakage compared to the planar reference, which might result from induced morphology changes. However, Table 1 and Figure 2a do not show an obvious correlation between block height~$H$ and the strength of the leakage. In addition, we observe reduced current densities of structured samples at voltages $V > 2.5$\,V. The reason for this might be hindered charge carrier injection at the blocks, and, as a result, a smaller area for the injection. Nevertheless, all current density-voltage-luminance (\textit{jVL}) curves show nearly the same exponential regime between \SI{2}{\volt} and  \SI{2.5}{\volt}, which suggests that the recombination process is undisturbed by the blocks \cite{fischer}. The luminance at high voltages drops similarly to the current density for structured samples. To reach higher efficiencies, one would expect a higher luminance for structured devices, but note that the EQE is proportional to the ratio between luminance and current density \cite{kohlerElectronicProcessesOrganic2015}. Here, the efficiency is enhanced because the current density drop is stronger than the luminance drop. For example, at  \SI{4}{\volt} the current density for sample G2 drops almost by a factor of three, while the luminance is only two times smaller. This leads to higher EQEs of internally structured OLEDs for all block parameters, which is further discussed in the following.

To further explain the nature of the EQE increase, the angular dependent spectra are depicted in Figure 3 for the planar reference OLED and structured sample G4. The normalized intensity is plotted as a function of corresponding photon energies and the in-plane wave vector~$k_\parallel$. The contour plot on the left represents a planar reference OLED and the one on the right shows an OLED with a 2D block array with a period of  \SI{1}{\micro\meter}. The structured OLED exhibits periodic sharp linear features in addition to the planar emission spectrum. These features can be identified as Bragg scattering modes of different orders~\cite{Fuchs:13}. This means that the EQE is increased by optical outcoupling and not due to a change in electrical performance. The angular resolved spectra of all OLEDs with TiO\textsubscript{2} blocks are given in the Supplementary Figure~1.   

\begin{figure}[h!]
	\centering
	\includegraphics[scale = 0.55]{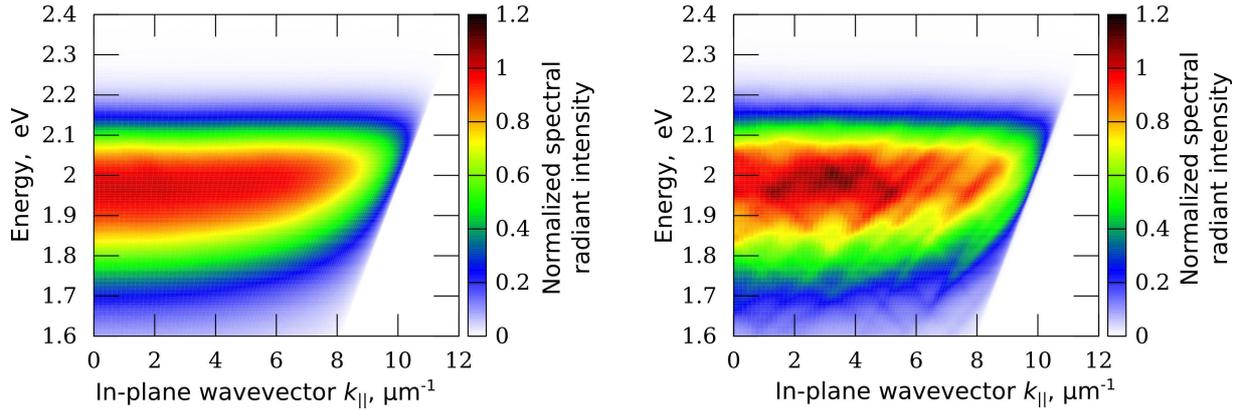}
	\caption{Angular resolved emission spectra of a) planar reference device and b) device with 2D block array with the period of \SI{1}{\micro\meter}.}
	\label{fig:gonio}
\end{figure}

The EQE over current density is plotted for planar reference and structured devices in Figure 2b. Additionally, this plot includes the EQE values for the same samples with an attached half-sphere that serves as an external light outcoupling structure. All the devices with internal outcoupling structures show an efficiency increase compared to the planar device independently of the period of the block arrays. Noticeably, the enhancement in the EQE is more pronounced when the half-sphere is applied. This indicates that some of the waveguided modes, which are scattered by the 2D block arrays, might still be coupled to the substrate modes. The latter are then extracted by the attached half-sphere.  

It is noteworthy that the leakage current which we see in electrical characteristics (Figure 2a) reduces the EQE in Figure 2b at low current densities. As a result, the EQE maxima are not at the same current densities. This complicates the comparison of OLEDs with varied block parameters. To exclude unwanted impact of the leakage current on the EQE, values will be taken at \SI[per-mode=symbol]{15}{\milli\ampere\per\square\centi\meter} in the following analysis. Nevertheless, the maximum EQE values for each OLED are greater than the values at \SI[per-mode=symbol]{15}{\milli\ampere\per\square\centi\meter}, e.g. the highest achieved EQE is \SI{23.6}{\%} for the sample G2 with internal structures and \SI{45.2}{\%} for the sample G7 with both internal and external structures.

\section{Discussion}

A desired goal for OLED research is to find the shapes and dimensions of light outcoupling structures that lead to the highest light output. 
Previously published simulations predict that there are ideal parameters of periodic scattering structures for light outcoupling in OLEDs. For example, R. Pfeifer \textit{et al.} \cite{6068687} suggest that gratings with a period of $\Lambda=800$\,nm and height of $H=60$\,nm are the most effective for achieving the maximal EQE of green OLEDs. Another study \cite{yueEnhancingOutCouplingEfficiency2012a} shows simulations for photonic crystals embedded in OLEDs and states that the optimal period is 800 -- 1000\,nm and height is 300\,nm. These two parameters are also calculated for wrinkled substrates by J. Park \textit{et al.} \cite{7416153} for OLEDs with the emission wavelenght around 530\,nm, where the most effective periods are between 800 and 1200\,nm  and heights between  100 and 200\,nm.

In this experimental study, we aim to find ideal block array parameters for a red bottom-emitting OLED. Therefore, we search for correlations between the EQE and block array parameters, which may help to predict the most effective light outcoupling structure. The 2D block array has two main spatial parameters: the period $\Lambda$, which is equal to the sum of $A$ and $B$, and the height $H$. Additionally, the spectrally resolved transmission through the substrate and the block arrays was measured for each sample and weighted with the emission spectrum of the OLED (see Supplementary Figure 2). As the transmission needs to be sufficiently high for bottom-emitting OLEDs, it is taken as one of the parameters as well. In total we use four main parameters for the evaluation: period $\Lambda$, ratio between $A$ and $B$, height $H$, and transmission $T$. However, since all the parameters are simultaneously varied, it becomes difficult to untangle the influence of each parameter.

To analyze the data we made scatter plots where it is possible to incorporate three variables in one plot. The evaluation is done by regression of the different parameters to the EQE while displaying a second parameter which is varied by systematic combination. This helps to visualize the relationship between the EQE and the two parameters at the same time by encoding their values in horizontal and vertical axes along with sizes of the points. 

The period $\Lambda$ is one of the key parameters for maximizing the OLED efficiency. For this reason, the EQE is plotted firstly over the period $\Lambda$ for OLEDs with internal and with both internal and external structures in Figure 4. As already mentioned in the results section, all the EQE values are taken at a higher current density of \SI[per-mode=symbol]{15}{\milli\ampere\per\square\centi\meter} to exclude the impact of electrical influence. In this scatter plot, the size of the points scales inversely with the $A/B$ ratio, which means the smaller the point, the higher the $A/B$ ratio and the smaller the distance between the blocks. 

Figure 4 shows that for internal structures only (green), the EQE is approximately constant over the period. However, for the OLEDs with the glass half-sphere attached (blue), the EQE seems to have a maximum at around $\Lambda~\approx$~1000 -- 1500\,nm. Interestingly, the OLED with the period of 1200\,nm deviates from the general trend, but for this particular grating, the $A/B$ ratio is much larger compared to neighboring samples. The point scaling is chosen to emphasize that this sample could be neglected. We can only speculate whether there is a physical reason or if this is statistical noise. However, the distance between the blocks $B$ in this sample is the smallest with $B = 200$\,nm (see Table 1), so the grating might represent only a weak perturbation for the photons since the subsequent layer deposition will smoothen the gap even more. As a result, the scattering is weaker. The same trend can be seen for two samples with identical periods of 1500\,nm. The one with the higher $A/B$ has a smaller EQE. It can be concluded that besides the ideal period $\Lambda~\approx$~1000 -- 1500\,nm, a small $A/B$ ratio with $A/B~\approx~1$ is also needed for a high EQE.

\begin{figure}[h!]
	\centering
	\includegraphics[scale = 0.55]{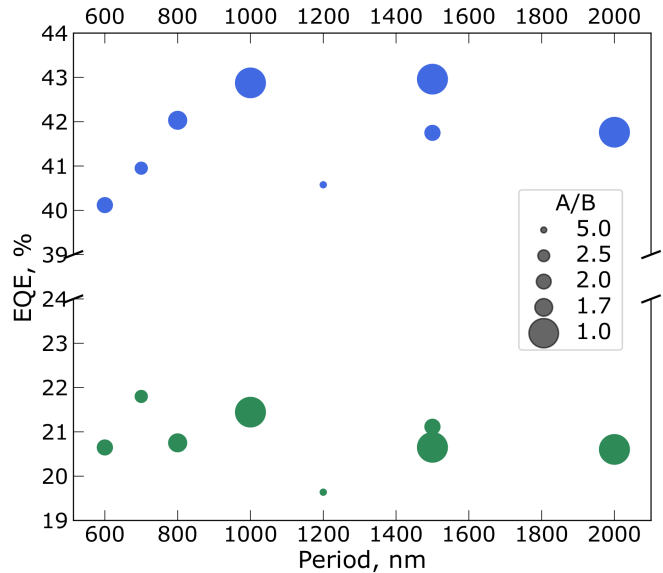}
	\caption{EQE values are plotted versus the period of the block arrays. Green points represent the samples with only internal structures and blue points the samples with internal and external structures. The EQE values are obtained at a current density of $j = 15$ mA/cm\textsuperscript{2}.}
	\label{fig:EQEs-period}
\end{figure}

As we can see, the impact on the EQE of the internal structures is more pronounced with the attached half-sphere. For this reason, the following evaluation is done only for samples with both internal and external structures.
Thus, all the variations of the parameters are presented in the matrix of plots in Figure 5. Here, EQE values always remain in the vertical axis in each plot. The horizontal axis is varied for each row, so that, e.g, plots a, b and c have the same x and y axes. Finally, the size of the points is varied with the parameters in each row. With this, we have all possible combination of the four parameters: the period $\Lambda$, the ratio $A/B$, the height $H$, and the transmission $T$. Additionally, the Pearson correlation coefficient $r$ is calculated for the parameters on the x- and y-axis, which has values between $r = -1$ and $r=1$ and quantifies whether a linear correlation of the EQE and a specific parameter is present \cite{edwardsIntroductionLinearRegression1976}. The value of 1 or -1 represents positive or negative correlation respectively. If $r=0$ then no linear correlation  is present.

\begin{figure}[h!]
	\centering
	\includegraphics[scale = 0.6]{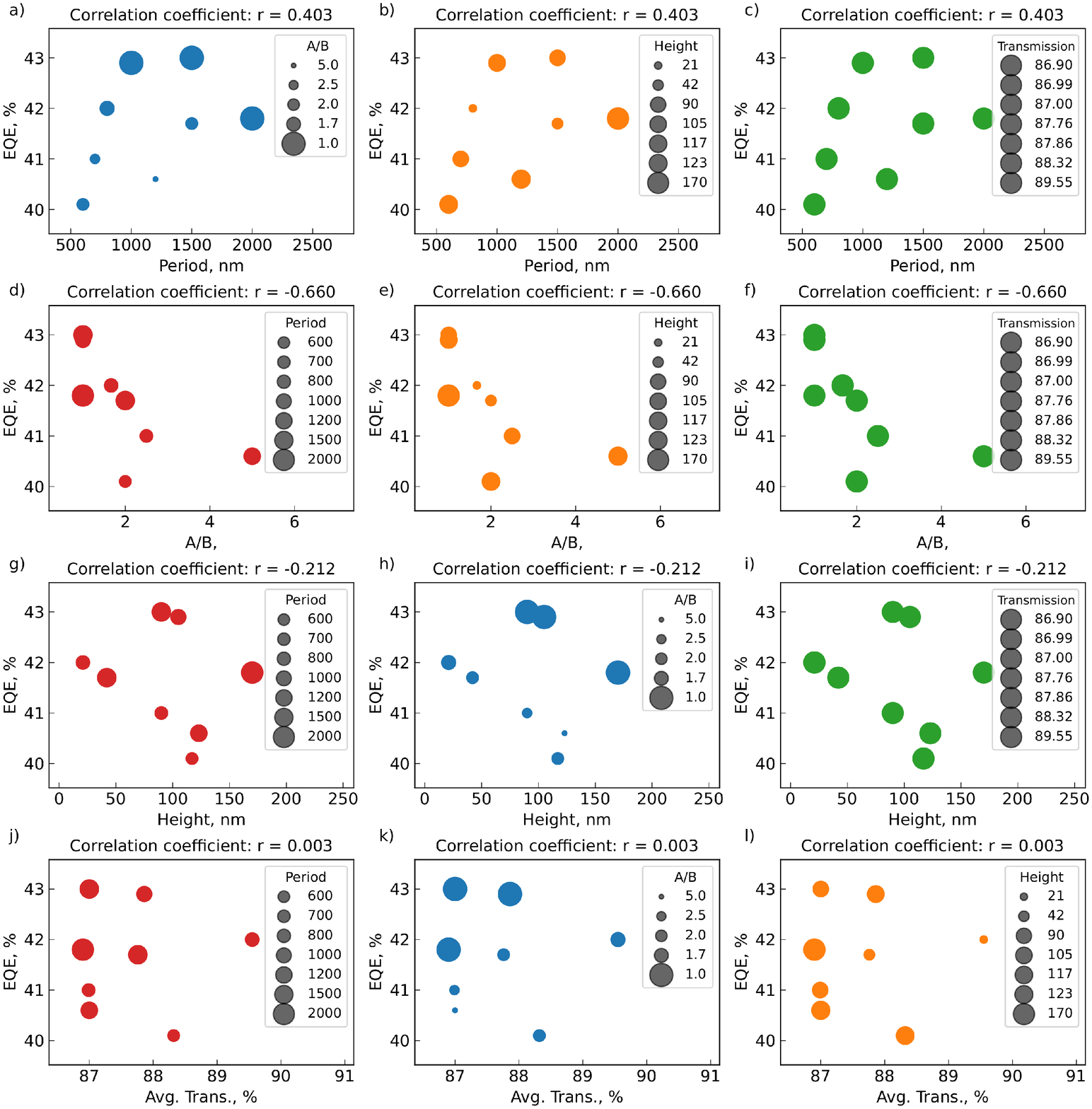}
	\centering
	\caption{Correlation between the EQEs and combinations of all four parameters of the blocks: period $\Lambda$, height $H$, ratio between block width and distance between the blocks $A/B$ and transmission $T$. The Pearsion correlation coefficient $r$ is calculated on the x- and y-axis parameters.}
	\label{fig:scatter}
\end{figure}

\enlargethispage{-1\baselineskip}
The EQE and the period $\Lambda$ have a modest linear correlation between each other with $r=0.403$ (Subfigures a, b and c). However, from Figure 4 we already expected a non-linear correlation with an optimal period of about 1000 -- 1500\,nm, which is in agreement with simulations from literature \cite{6068687,yueEnhancingOutCouplingEfficiency2012a, 7416153}. As we have seen before, the $A/B$ ratio influences the efficiency as well. Remarkably, the EQE and $A/B$ ratio exhibit the strongest linear correlation with a coefficient of $r=-0.660$ (Subfigures d, e and f). The lowest correlation is found between the EQE and transmission with $r=0.003$ (Subfigures j, k and l). This emphasizes that a high transmission does not necessarily lead to a high EQE \cite{hoangOptimumThicknessEpsilon2017, fuchsQuantitativeAllocationBragg2013}. The height does not show a strong correlation to the EQE as well. Even though height and transmission do not significantly influence the EQE, they still exhibit a correlation between each other with a coefficient of $r=-0.654$ (Supplementary Figure 3), which seems logical as increasing TiO\textsubscript{2} height will strengthen the reflection \cite{ZHAO200779}. Further correlations with all parameters and even parameter combinations have been calculated systematically. However, not all of them were meaningful, e.g. $A$ versus $A/B$. The more reasonable combinations are provided in Supplementary Figure 3 without further discussion, as it is unclear whether there is a causality. 

\section{Conclusion}

In this paper, we investigated 2D TiO\textsubscript{2} block arrays as internal light outcoupling structures in bottom-emitting OLEDs. We varied the block parameters in order to identify the parameter values that achieve the highest possible EQE. The 2D TiO\textsubscript{2} block arrays were successfully implemented beneath the OLED layer sequence. Although the blocks slightly increase the leakage current of the OLEDs, all of devices showed an EQE increase. The highest EQE achieved for the OLED with only internal structures is \SI{23.6}{\%} with the period $\Lambda = 700$\,nm and is \SI{45.2}{\%} at $\Lambda = 1500$\,nm for both internal and external structures. Note that the planar reference OLED with a half-sphere attached reaches at best an EQE of \SI{37.7}{\%}, which indicates that a fraction of the waveguided modes were scattered in the substrate, which is then extracted by the half-sphere.

Using scatter plots as a visualization method, we investigated the complex correlation between block parameters and the EQE. Despite the multi-parameter problem emphasized before, we suspect the ideal period of 2D block arrays for maximum light outcoupling to be between 1000\,nm and 1500\,nm, which is in a good agreement with previously published works. Remarkably, the block width-to-distance ratio $A/B$ must also be considered for achieving high EQEs. It should be close to unity, or possibly even smaller, so that the distance between the blocks $B$ is similar in size to the blocks~$A$. This result suggests that too narrow of a distance between blocks might not provide sufficient perturbation for scattering of the photons because the layers deposited on top smooth the corrugation.  
Furthermore, for the investigated parameter range, the EQE did not depend on either the block height nor on the transmission.

The data obtained during this work gives an insight into the parameter dependencies for further simulations and investigations of light outcoupling structures to achieve high efficiencies in OLEDs. Further experimental investigations can be done to understand and ultimately optimize the light scattering of the blocks, i.e. locally and spectrally resolved imaging of OLEDs during emission. Finally, OLEDs with different emission spectra should be studied to further strengthen the understanding of the complex optics of such devices. 

\section*{Data availability}
The data that support the findings of this study are available from the corresponding author upon reasonable request.

\printbibliography

\newpage

\section{Supplementary Information}

\setcounter{figure}{0} 
\renewcommand{\figurename}{Supplementary Figure}

\begin{figure}[h]
	\includegraphics[scale = 1]{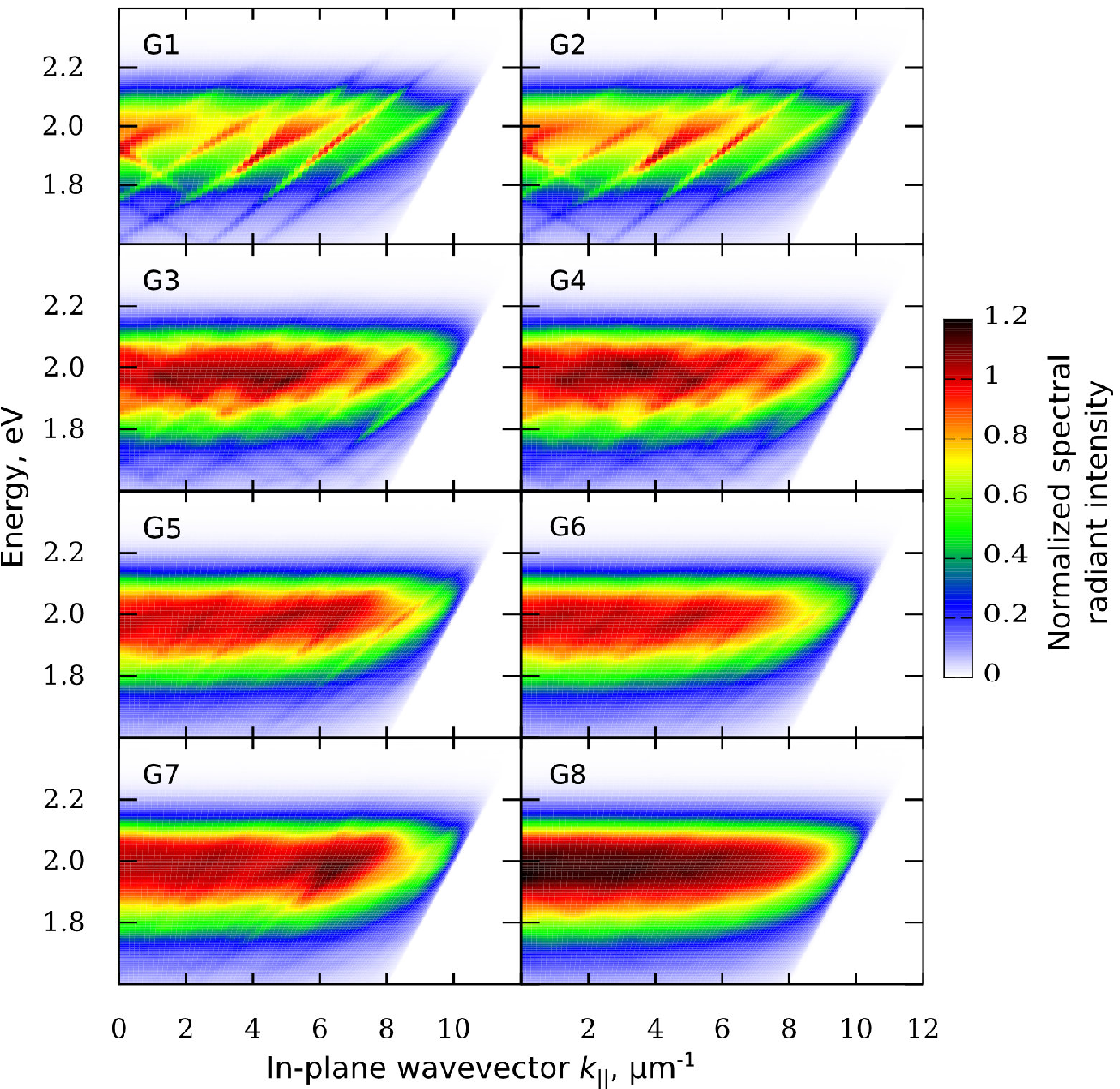}
	\centering
	\caption{Normalized spectral intensities of samples G1-G8 plotted over photon energies and in-plane wave vector $k_\parallel$. The intensity is normalized for each sample individually.}
	\label{fig:ref-angles}
\end{figure}

\begin{figure}[h]
	\includegraphics[scale = 0.5]{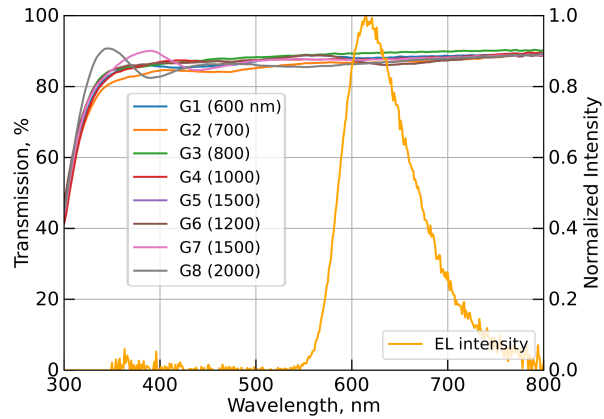}
	\centering
	\caption{Spectrally resolved transmission of selected block arrays and electro-luminescent spectrum that is used to calculate the average transmission.}
	\label{fig:transmission}
\end{figure}

\begin{figure}[h]
	\includegraphics[scale = 0.7]{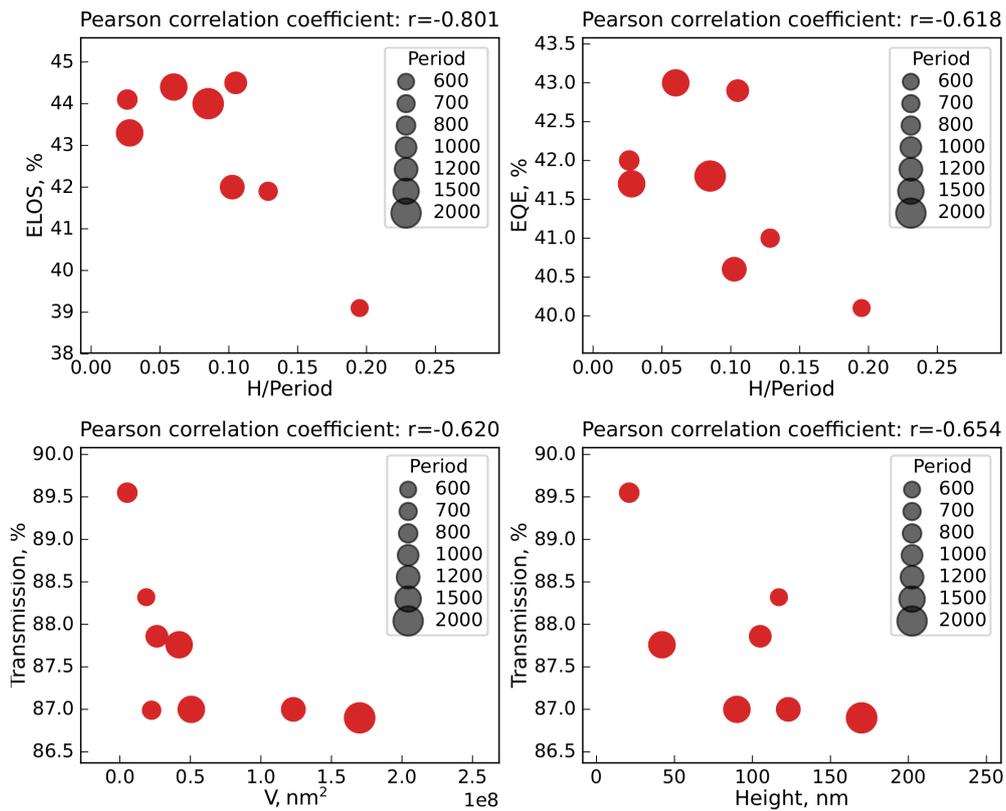}
	\centering
	\caption{Additional scatter plots with comparably high linear correlation. The abbreviation ELOS is an alternative efficiency to evaluate the effectiveness of light outcoupling structures~\cite{will2019efficiency}. The symbol $V$ represents the block volume with $V = A^2 \cdot H$. }
	\label{fig:scatter-SI}
\end{figure}

\end{document}